\documentclass[twocolumn,pra,10pt,showpacs]{revtex4}
\usepackage{graphicx,amsmath,amssymb}

\newcommand{\beqa}{\begin{eqnarray}}
\newcommand{\eeqa}{\end{eqnarray}}
\newcommand{\ket}[1]{| #1 \rangle}
\newcommand{\bra}[1]{\langle #1 |}

\newcommand{\opua}{\hat{U}_A}
\newcommand{\opub}{\hat{U}_B}

\newcommand{\density}{\hat{\rho}}

\newcommand{\g}{\Gamma}

\newcommand{\gmax}{\Gamma_{{\rm \mbox{sup}}}}

\newcommand{\phase}[1]{\varphi_{\rm #1} }
\newcommand{\oprho}{\hat{\rho}}

\begin{document}


\title{Entanglement criterion for pure $M\otimes N$ bipartite quantum states}

\author{Hoshang Heydari}
\email{hoshang@imit.kth.se} \homepage{http://www.ele.kth.se/QEO/}
\affiliation{Department of Microelectronics and Information
Technology, Royal Institute of Technology (KTH), Electrum 229,
SE-164 40 Kista, Sweden}

\author{Gunnar Bj\"{o}rk}
\affiliation{Department of Microelectronics and Information
Technology, Royal Institute of Technology (KTH), Electrum 229,
SE-164 40 Kista, Sweden}

\date{\today}

\begin{abstract}
We propose a  entanglement measure for pure $M \otimes N$
bipartite quantum states. We obtain the measure by generalizing
the equivalent measure for a $2 \otimes 2$ system, via a $2
\otimes 3$ system, to the general bipartite case. The measure
emphasizes the role Bell states have, both for forming the
measure, and for experimentally measuring the entanglement. The
form of the measure is similar to generalized concurrence. In the
case of $2 \otimes 3$ systems, we prove that our measure, that is
directly measurable, equals the concurrence. It is also shown that
in order to measure the entanglement, it is sufficient to measure
the projections of the state onto a maximum of $M(M-1)N(N-1)/2$
Bell states.
\end{abstract}

\pacs{03.67.Mn,42.50.Dv,42.50.Hz,42.65.Ky}

\maketitle

\section{Introduction}

The concept of quantum entanglement is not new, it goes back to
the early days of quantum theory where it was initiated by
Einstein, Podolsky, and Rosen \cite{EPR35}, and Schr\"{o}dinger
\cite{Sch35}. Many years has passed since the dawn of quantum
mechanics, but we have still not been able to solve the enigma of
entanglement, e.g., finding a complete mathematical model to
describe and quantify this interesting feature of quantum
mechanical systems, and in the same time reveal the physical
implications of this feature. During recent years, separability
and entanglement has been a vital research field. Peres pioneered
quantification of entanglement by showing that a necessary
criterion for separability was positivity of the density matrix
upon partial transposition \cite{Peres96}. Soon thereafter,
Horodeckis proved that the criterion was also sufficient
\cite{Hor96}.

Recently, several quantitative measures of entanglement for
nonseparable states, such as entanglement of formation
\cite{Bennett96}, distillable entanglement \cite{Bennett962},
relative entropy of entanglement \cite{Vedral97}, concurrence
\cite{Wootters98}, or concurrence related measures
\cite{Audenaert,Rungta01,Albeverio,Gerjuoy,Cereceda} have been
suggested. In particular, the definition of concurrence is based
on the spin-flip operation, and Rungta et al. \cite{Rungta01} have
generalized this operator and defined I-concurrence for pure
bipartite state in any dimension. For the mixed $2\otimes M$
bipartite state we have a separability criterion given in
\cite{Kraus00} and a lower bound of concurrence of mixed such
quantum states \cite{Lozinski03}. Entanglement witnesses is
another method of detecting entanglement \cite{Lewen00}.

In this paper, we develop a measure for general pure $N \otimes M$
bipartite quantum states, inspired by the measure of entanglement
proposed in \cite{Soto02,Hosh}. It is based on bipartite phase
sums and differences, and we conjecture that if it is properly
normalized, it equals the I-concurrence for all pure bipartite
states \cite{Rungta01}. In Sec. \ref{sec:2x2} we briefly discuss
the physical grounds for our measure for $2 \otimes 2$ systems
(two qubits). In Sec. \ref{sec:2x3} we argue what a similar
measure should be for a $2 \otimes 3$. In Sec. \ref{sec: math}, we
derive a measure using a relative-phase positive operator valued
measure (POVM) as the entanglement quantifier. In Sec.
\ref{sec:comparison} we compare our measurement to existing
measures. In Sec. \ref{sec:generalization} we extent the measure
to encompass bipartite systems of any dimension. In Sec.
\ref{sec:examples} we apply our measure to a few sample states and
discuss its normalization and the role of the Bell states.
Finally, in Sec. \ref{sec:conclusions} we summarize our findings.
The main novelty with our paper is not the measure itself, since
it is essentially proportional to I-concurrence, but rather the
way the measure is derived. We have based our derivation on
physical argumentation in contrast to earlier derivations that are
primarily based on mathematical arguments.

\section{Entanglementbetween two qubits, a brief review}
\label{sec:2x2}

In an earlier paper we investigated the entanglement properties of
a $2 \otimes 2$ bipartite state \cite{Hosh}. The starting point of
that investigation was the assumption that the entanglement
properties are, or can be, expressed in the bipartite state's
joint phase properties. On basis of this assumption, we found that
the state's non-local properties are found in two of the state's
off-diagonal density matrix coefficients (when the density matrix
was expressed in the standard basis $\ket{11}$, $\ket{12}$,
$\ket{21}$, and $\ket{22}$). Examining the reasons for this, a
rather simple physical reasoning shows why this is the case.
Tensor multiplying two qubit density matrices, $\oprho =
\oprho_{\rm A} \otimes \oprho_{\rm B}$, where each factor
$\oprho_i$,  $i = {\rm A}, {\rm B}$, is of the form
\begin{equation}
\oprho_i =  \left (
\begin{array}{cc}
\rho_{11}  & |\rho_{12}| e^{i \varphi_i} \nonumber \\
|\rho_{12}| e^{-i \varphi_i} & 1-\rho_{11}
\end{array}
\right ) ,
\end{equation}
and subdividing the ensuing $4 \times 4$ matrix into four $2
\times 2$ quadrants, it is clear that the upper left quadrant does
not contain any information about qubit A's phase $\varphi_{\rm
A}$. The lower right quadrant lacks this information, too. Due to
hermiticity, the upper right and the lower left quadrant contain
the same information. Therefore it suffices to consider, e.g., the
upper right quadrant. Of its four coefficients, its diagonal terms
(that is, the coefficients 13 and 24 of the joint system density
matrix) do not contain information about qubit B's phase
$\varphi_{\rm B}$. Hence, all the joint phase information is
collected in the off-diagonal coefficients 14 and 23, and we found
that a relevant measure of entanglement for a two qubit system was
\begin{equation}
\gmax = {\rm Sup} ( 2 ||\rho_{14}|-|\rho_{23}|| ) ,
\end{equation}
where Sup refers to the supremum of the function with respect to
any local unitary transformation(s).

One then notes that by a local phase rotations, it is always
possible to make $\rho_{14}$ and $\rho_{23}$ simultaneously real.
In this case
\begin{equation}
{\rm Re}(\rho_{14}) = |\rho_{14}| =
\bra{\Psi_{+}}\oprho\ket{\Psi_{+}}-\bra{\Psi_{-}}\oprho\ket{\Psi_{-}},
\end{equation}
\begin{equation}
{\rm Re}(\rho_{23}) = |\rho_{23}| =
\bra{\Phi_{+}}\oprho\ket{\Phi_{+}}-\bra{\Phi_{-}}\oprho\ket{\Phi_{-}},
\end{equation}
where $\ket{\Psi_{\pm}} = (\ket{11} \pm \ket{22})/\sqrt{2}$,
$\ket{\Phi_{\pm}} = (\ket{12} \pm \ket{21})/\sqrt{2}$, and Re
denotes the real part. That is, the entanglement is simply the
maximum of the difference between the state's projections onto the
Bell states. Hence, it can be measured as the Bell-analyzer
visibility \cite{Hosh}.

\section{An entanglement measure for a $2 \times 3$ bipartite state}
\label{sec:2x3}

Now we apply a physical reasoning similar to that in Sec.
\ref{sec:2x2} to a bipartite $2 \otimes 3$ system (a qubit and a
qutrit). We start by subdividing the system density operator (a $6
\times 6$ matrix) into four $3 \times 3$ matrix quadrants. It is
clear that the upper left and lower right quadrants do not contain
information about the qubit phase $\varphi_{\textrm{A}}$. The
remaining two quadrants contain the same information so let us
focus on, e.g., the upper right quadrant. Disregarding the
diagonal coefficients in this quadrant (they contain information
about $\varphi_{\textrm{A}}$ but not about the qutrit phases), we
see that the joint phase properties of the state lies in the
density matrix coefficients 15, 16, 24, 26, 34, and 35. These
coefficients correspond to the projectors $\ket{11}\bra{22}$,
$\ket{11}\bra{23}$, $\ket{12}\bra{21}$, $\ket{12}\bra{23}$,
$\ket{13}\bra{21}$, and $\ket{13}\bra{22}$, respectively. Consider
now the two possible complete Bell bases:
$$\ket{\Psi_{1}}=\frac{1}{\sqrt{2}}(\ket{11}+\ket{22}),~\ket{\Psi_{2}}=\frac{1}{\sqrt{2}}(\ket{11}-\ket{22}),$$
$$\ket{\Psi_{3}}=\frac{1}{\sqrt{2}}(\ket{12}+\ket{23}),~\ket{\Psi_{4}}=\frac{1}{\sqrt{2}}(\ket{12}-\ket{23}),$$
and
$$\ket{\Psi_{5}}=\frac{1}{\sqrt{2}}(\ket{13}+\ket{21}),~\ket{\Psi_{6}}=\frac{1}{\sqrt{2}}(\ket{13}-\ket{21}),$$
or
$$\ket{\Phi_{1}}=\frac{1}{\sqrt{2}}(\ket{11}+\ket{23}),~\ket{\Phi_{2}}=\frac{1}{\sqrt{2}}(\ket{11}-\ket{23}),$$
$$\ket{\Phi_{3}}=\frac{1}{\sqrt{2}}(\ket{12}+\ket{21}),~\ket{\Phi_{4}}=\frac{1}{\sqrt{2}}(\ket{12}-\ket{21}),$$
and
$$\ket{\Phi_{5}}=\frac{1}{\sqrt{2}}(\ket{13}+\ket{22}),~\ket{\Phi_{6}}=\frac{1}{\sqrt{2}}(\ket{13}-\ket{22}).$$
The two bases can be obtained from each other by permutation of
any two of the qutrit states. However, such a permutation changes
all six states, that is, one whole basis is transformed into the
other. We note that
\begin{equation}
\rho_{15} + \rho_{51} = 2 {\rm Re} (\rho_{15}) =
\bra{\Psi_{1}}\oprho\ket{\Psi_{1}}-\bra{\Psi_{2}}\oprho\ket{\Psi_{2}},
\end{equation}
and
\begin{equation}
\rho_{16} + \rho_{61} = 2 {\rm Re} (\rho_{16}) = \bra{\Phi_{1}}
\oprho \ket{\Phi_{1}} + \bra{\Phi_{2}} \oprho \ket{\Phi_{2}} ,
\end{equation}
etc. By applying phase shifts, local to the qutrit B, it is always
possible to make all three coefficients $\rho_{15}$, $\rho_{26}$,
and $\rho_{34}$ real, simultaneously. Hence, the absolute values
of density matrix coefficients 15, 26, and 34 are associated with,
and can be obtained from, a Bell-state analysis using the complete
basis set $\{\ket{\Psi_{i}}\}$, while, in a similar manner,
coefficients 16, 24, and 35 are associated with the complete and
noncompatible set $\{\ket{\Phi_{i}}\}$. Intuitively, one would
expect the entanglement to be greatest if only the states in one
of the sets were excited. Moreover, the entanglement should be
maximized if only one of the states in one of the sets were
excited. A reasonable measure of entanglement would therefore be
\begin{eqnarray}\label{eq:
central equation} \g &=& ({\cal N}_2
(||\rho_{15}|-|\rho_{24}||^{2}+||\rho_{26}|-|\rho_{35}||^{2}\\\nonumber
&&+||\rho_{34}|-|\rho_{16}||^{2}))^{\frac{1}{2}},
\end{eqnarray}
where, again, $\g$ is not invariant to transformations local to
qubit A and qutrit B, respectively, so the state's entanglement is
understood to be given by $\gmax$, the supremum of $\g$ taken over
all possible local unitary transformations. ${\cal{N}}_2$ is a
normalization factor that, if taken to be 2, makes $0 \leq \g \leq
1$. Eq. (\ref{eq: central equation}) is our central result for $2
\otimes 3$ systems. Due to the absolute signs, the expression is
symmetric with respect to the two Bell basis sets. It is also
clear that for any separable state, $\density =
\density_{\textrm{A}} \otimes \density_{\textrm{B}}$, one gets
\begin{equation}
\g=0
\end{equation}
since
$\rho_{15}=\rho_{\textrm{A}12}\rho_{\textrm{B}12}$,
$\rho_{24}=\rho_{\textrm{A}12}\rho_{\textrm{B}12}^\ast$,
$\rho_{26}=\rho_{\textrm{A}12}\rho_{\textrm{B}23}$,
$\rho_{35}=\rho_{\textrm{A}12}\rho_{\textrm{B}23}^\ast$, and
$\rho_{16}=\rho_{\textrm{A}12}\rho_{\textrm{B}13}$,
$\rho_{34}=\rho_{\textrm{A}12}\rho_{\textrm{B}13}^\ast$. These
relations explain why we have chosen the specific pairing between
the coefficients in sets $\{\rho_{15},\rho_{26},\rho_{34} \}$ and
$\{\rho_{16},\rho_{24},\rho_{35}\}$ associated with the respective
Bell basis.

\section{A mathematical derivation}
\label{sec: math}

In order to eventually generalize the results, we need to be on a
little bit firmer mathematical ground. We start by introducing the
basis vectors $\ket{1}_{\rm A}, \ket{2}_{\rm A},$ and
$\ket{1}_{\rm B}, \ket{2}_{\rm B}, \ket{3}_{\rm B}$ for subsystems
A and B. Subsequently we form the Hermitian operator
\begin{equation}\label{eq: equationrr}
 \hat{\Delta}_{\rm A }(\phase{A;12}) = \frac{1}{2
\pi}( \hat{I} + e^{i \phase{A;12}} \ket{1}_{\rm A}\bra{2}_{\rm A}
+ {\rm h.c.} ) ,
\end{equation} where h.c. denotes the hermitian conjugate. In the
same manner we define
\begin{eqnarray}
 \hat{\Delta}_{\rm B }(\phase{B;12},\phase{B;13},\phase{B;23})&=&\nonumber
\frac{1}{2\pi}( \hat{I} + e^{i \phase{B;12}} \ket{1}_{\rm
B}\bra{2}_{\rm B}
\\ && + e^{i \phase{B;13}} \ket{1}_{\rm B}\bra{3}_{\rm B}
\\\nonumber && + e^{i \phase{B;23}} \ket{2}_{\rm B}\bra{3}_{\rm B}
+ {\rm h.c.} ) .
\end{eqnarray}
The bipartite system's phase properties are described by the
operator
\begin{equation}
\hat{\Delta}(\phase{A;12},\phase{B;12},\phase{B;13},\phase{B;23})=
\hat{\Delta}_{\rm A } \otimes \hat{\Delta}_{\rm B } . \label{eq:
delta}
\end{equation}
We can re-express this operator in terms of the system's sum and
difference phases \begin{equation}\varphi^{k,l}_{p,q \pm} =
\phase{A;kl} \pm \phase{B;pq} . \label{eq: sums and
differences}\end{equation} The linear dependence between
$\phase{A;kl}$, $\phase{B;pq}$ and $\varphi^{k,l}_{p,q +}$,
$\varphi^{k,l}_{p,q -}$ allows us to express the operator in
(\ref{eq: delta}) as a function of the sum and difference phases.
Hence we can write $\hat{\Delta}(\varphi^{1,2}_{1,2
+},\varphi^{1,2}_{1,2 -},\varphi^{1,2}_{1,3 +},\varphi^{1,2}_{1,3
-},\varphi^{1,2}_{2,3 +},\varphi^{1,2}_{2,3 -})$. Next we compute
to what extent the density operator depends on these phase sums
and differences. Since the sum and difference phase POVM is a
periodic function of the phases, we can compute the Fourier
components of the POVM's expectation value. We define, e.g.,
\begin{eqnarray}
\Gamma^{1,2}_{1,2 +} &=&\nonumber  \frac{1}{2} |\int_{2 \pi}
d\varphi^{1,2}_{1,2 +} e^{i \varphi^{1,2}_{1,2 +}}\\\nonumber
&&{\rm Tr}(\oprho [ \hat{\Delta}(\varphi^{1,2}_{1,2
+},\varphi^{1,2}_{1,2 -},\varphi^{1,2}_{1,3 +} ,\varphi^{1,2}_{1,3
-},\varphi^{1,2}_{2,3 +},\varphi^{1,2}_{2,3 -})\\\nonumber &&  +
\hat{\Delta}(\varphi^{1,2}_{1,2 +}+\pi,\varphi^{1,2}_{1,2
-}+\pi,\varphi^{1,2}_{1,3 +}+\pi,\varphi^{1,2}_{1,3 -}+\pi\\
&&,\varphi^{1,2}_{2,3 +}+\pi,\varphi^{1,2}_{2,3 -}+\pi) ]) |
 . \label{eq:gammas}
\end{eqnarray}
The addition of all the $\pi$ terms in the right-most term inside
the trace operation above makes the functions $\Gamma^{k,l}_{p,q
+}$ above insensitive to the diagonal coefficients of the density
matrix, so only the joint (nonseparable) properties are probed by
$\Gamma^{k,l}_{p,q +}$. Finally, we compute
\begin{eqnarray}\label{eq: final expression}
  \Gamma & = & \nonumber ( 2\pi {\cal N}_2[ ||\Gamma^{1,2}_{1,2 +}|-|\Gamma^{1,2}_{1,2 -}||^2
  + ||\Gamma^{1,2}_{1,3 +}|-|\Gamma^{1,2}_{1,3 -}||^2 \\\nonumber &&
  + ||\Gamma^{1,2}_{2,3 +}|-|\Gamma^{1,2}_{2,3 -}||^2  ]
  )^{1/2} \nonumber \\
  & = &  ( {\cal N}_2[ ||\oprho_{15}|-|\oprho_{24}||^2
  + ||\oprho_{26}|-|\oprho_{35}||^2 \\\nonumber &&
  + ||\oprho_{34}|-|\oprho_{16}||^2]
  )^{1/2} .
\end{eqnarray}
The result is identical to the one we obtained by reasoning in Eq.
(\ref{eq: central equation}), above. The derivation may seem
rather \textit{ad hoc}, but follows our derivation for the similar
entanglement measure for $2 \otimes 2$ systems. The various steps
are motivated in \cite{Hosh}.

\section{A comparison to existing measures}
\label{sec:comparison}

In this section we compare our degree of entanglement with
concurrence for pure bipartite quantum systems. There are several
attempts to generalize concurrence for pure bipartite quantum
systems. One of these generalizations is called I-concurrence and is
defined in terms of a super operator that is a generalization of a
spin 1/2 flip-operation to higher Hilbert-space dimensions. For a
pure state $\ket{\Phi}$, of dimension $N\otimes M$, I-concurrence
is defined \cite{Rungta01} as
\begin{equation}
C_I=\sqrt{2(1-{\rm Tr}(\varrho^{2}_{\textrm{A}}))}=\sqrt{2(1-{\rm
Tr}(\varrho^{2}_{\textrm{B}}))} ,
\end{equation}
where $\varrho_{\textrm{A}}={\rm
Tr}_{\textrm{B}}(\ket{\Phi}\bra{\Phi})$ and
$\varrho_{\textrm{B}}={\rm
Tr}_{\textrm{A}}(\ket{\Phi}\bra{\Phi})$. Let $\ket{\Psi}$ be a
pure state for a  $2\otimes 3$ bipartite quantum systems given by
\begin{equation}
\ket{\Psi}=
\sum^{2}_{k=1}\sum^{3}_{l=1}\alpha_{kl}\ket{k}\otimes\ket{l} =
\sum^{2}_{k=1}\sum^{3}_{l=1}\alpha_{kl}\ket{kl}
\end{equation}
where $\{\ket{k}\}$ and $\{\ket{l}\}$ are two complete orthonormal
basis vector sets spanning a $M$- and $N$-dimensional Hilbert
space, respectively, and where
$\sum^{2}_{k=1}\sum^{3}_{l=1}|\alpha_{kl}|^{2} = 1$. For such a
state, the I-concurrence $C_I$ coincides with the concurrence $C$
for the state \cite{Gerjuoy}, and it can be rather simply
expressed in the state's probability amplitudes as
\begin{eqnarray}
C&=&\nonumber(2(|\alpha_{11}
\alpha_{22}-\alpha_{12}\alpha_{21}|^{2}+|\alpha_{11}
\alpha_{23}-\alpha_{13}\alpha_{21}|^{2}\\
& & + |\alpha_{12}
\alpha_{23}-\alpha_{13}\alpha_{22}|^{2}))^{\frac{1}{2}} .
\label{eq: two by three concurrency}
\end{eqnarray}
(Very similar expressions have been also derived in
\cite{Albeverio} and in \cite{Cereceda}). This expression should
be compared to our entanglement measure
\begin{eqnarray}
\gmax &=&\nonumber {\rm Sup}[{\cal N}_2(|||\alpha_{11}
\alpha^{*}_{22}|-|\alpha_{12}\alpha^{*}_{21}||^{2}\\\nonumber &&
+||\alpha_{11}
\alpha^{*}_{23}|-|\alpha_{13}\alpha^{*}_{21}||^{2}\\ && +
||\alpha_{12}
\alpha^{*}_{23}|-|\alpha_{13}\alpha^{*}_{22}||^{2}|)]^{\frac{1}{2}}
. \label{eq: gamma}
\end{eqnarray}
It is seen that for ${\cal N}_2=2$, we get $\gmax \leq C$. Suppose we
rotate this state with local unitary operations $\opua \otimes
\opub$ such that
\begin{equation}
\opua \otimes \opub \ket{\psi} =
\sum^{2}_{k=1}\sum^{3}_{l=1}\alpha'_{kl}\ket{kl} .
\end{equation}
The state's concurrence is invariant with respect of the
operation. Hence, if it is always possible to make, e.g.
$\alpha'_{11}=\alpha'_{23}=0$, then the two expressions above
coincide to yield
$(2(|\alpha'_{12}\alpha'_{21}|^{2}+|\alpha'_{13}\alpha'_{21}|^{2}+
|\alpha'_{13}\alpha'_{22}|^{2}))^{\frac{1}{2}}$, and such a
transformation maximizes $\g$. We shall now prove that this is
always possible. Consider the local unitary transformations
\begin{equation}
\opua(\theta, \phi) = \left (
\begin{array}{cc}
\cos\theta & i e^{i \phi} \sin\theta \nonumber \\
i  \sin\theta & e^{i \phi} \cos\theta
\end{array}
\right )\end{equation} and
\begin{equation}
\opub(\vartheta, \varphi) = \left (
\begin{array}{ccc}
1 & 0 & 0 \nonumber \\
0 & \cos\vartheta & i e^{i \varphi} \sin\vartheta \nonumber \\
0 & i  \sin\vartheta & e^{i \varphi} \cos\vartheta
\end{array}
\right ) .\end{equation} From these transformations we get
\begin{equation}
\alpha'_{11} = \alpha_{11} \cos\theta + i \alpha_{21} e^{i \phi}
\sin\theta.
\end{equation}
It is seen that it is always possible to make $\alpha'_{11}=0$ by
an appropriate choice of $\theta$ and $\phi$. We also get
\begin{eqnarray}
\alpha'_{23} & = & i  \sin\vartheta (i  \alpha_{12}\sin\theta +
\alpha_{22} e^{i \phi} \cos\theta)) \nonumber \\
& & + e^{i \varphi}\cos\vartheta (i  \alpha_{13} \sin\theta +
\alpha_{23} e^{i \phi} \cos\theta) .
\end{eqnarray}
For a given choice of $\theta$ and $\phi$, the expressions in the
parenthesis on the right hand side of the equation above are two
fixed complex numbers Therefore, it is always simultaneously
possible to assure that $\alpha'_{23}=0$ by an appropriate choice
of the parameters $\vartheta$ and $\varphi$. Hence, the proof that
our measure coincides with the concurrence for pure states is
completed.

From (\ref{eq: gamma}) above, we see that it is formally also
possible to prove that the measures coincide if, e.g.,
$\alpha'_{11}=\alpha'_{23}=0$, or if
$\alpha'_{11}=\alpha'_{22}=0$, $\alpha'_{11}=\alpha'_{12}=0$, or
$\alpha'_{11}=\alpha'_{13}=0$, where we have looked at the
possibilities where $\alpha'_{11}=0$. (The other possibilities
lead to equivalent conclusions.) The first case was already
proven, above. In the second case, a proof outlined like the one
pertaining to the first case, shown above, can be used. In the
third case we can prove the assertion as follows: The pure state
can be written $\ket{\psi}= \ket{1} \otimes (\alpha_{11} \ket{1} +
\alpha_{12} \ket{2} + \alpha_{13} \ket{3}) + \ldots$. We see that
in order to prove our assertions, it only makes sense to consider
rotations in subspace B. However, in order to make
$\alpha'_{11}=\alpha'_{12}=0$ we need to find three new basis
vectors in the subspace that are mutually orthogonal, and that
makes $\alpha_{11} \ket{1} + \alpha_{12} \ket{2} + \alpha_{13}
\ket{3} \rightarrow \alpha'_{13} \ket{3'}$. It is quite obvious
that the choice $\ket{3'} = (\alpha_{11} \ket{1} + \alpha_{12}
\ket{2} + \alpha_{13} \ket{3})/\sqrt{\cal{N}}$, where $\cal{N}$ is
a normalization factor, and where $\ket{1'}$ and $\ket{2'}$ can be
chosen arbitrary, as long as they are orthogonal to $\ket{3'}$ and
to each other, satisfies our requirement. In the fourth case we
have $\ket{\psi}= (\alpha_{11} \ket{1} + \alpha_{21} \ket{2})
\otimes \ket{1} + \ldots$. We see that in order to get
$\alpha'_{11}=\alpha'_{21}=0$, we need to find a unitary
transformation in space A rendering $\alpha_{11} \ket{1} +
\alpha_{21} \ket{2} \rightarrow 0 \ket{1'} + 0 \ket{2'}$. In is
obvious that if either $\alpha_{11}$ or $\alpha_{21}$ are nonzero,
no such transformation can be found.

As a consequence of the conclusions above, we can deduce that in
order to measure the entanglement of a pure state, one can do Bell
measurements, but it is necessary to use not one, but a minimum of
two different Bell bases. Suppose,
e.g., that we have found a basis such that
$\alpha'_{11}=\alpha'_{23}=0$. We then need to find
$|\rho'_{24}|$, $|\rho'_{35}|$, and $|\rho'_{34}|$ to get $\gmax$.
The first of the density matrix coefficients can be found by
projection on the states $\ket{\Phi_{1}}$ and $\ket{\Phi_{2}}$
subtracting one of the outcome probabilities from the other,
taking the absolute of the difference, and dividing by two. To
obtain $\rho'_{35}$, we can likewise project on the states
$\ket{\Phi_{5}}$ and $\ket{\Phi_{6}}$ (orthogonal to the first
pair of projectors). Finally, to get $\rho'_{34}$, we need to
project on the states $\ket{\Psi_{5}}$ and $\ket{\Psi_{6}}$. These
two states, however, are not orthogonal to the any of the previous
four Bell states.



\section{Generalizing to $M \otimes N$ bipartite quantum states}
\label{sec:generalization}

The derivation of the entanglement measure for a $2 \otimes 3$
system made in Sec. \ref{sec: math} can be extended to an
arbitrary bipartite system. We label the basis vectors
$\ket{1}_{\rm A}, \ldots , \ket{M}_{\rm A},$ and $\ket{1}_{\rm B},
\ldots, \ket{N}_{\rm B}$. Subsequently we form the operator
\begin{equation} \hat{\Delta}_{\rm A } = \frac{1}{2\pi}( \hat{I} + \sum_{k=1}^{M-1} \sum_{l=k+1}^{M} e^{i
\phase{A;kl}} \ket{k}_{\rm A}\bra{l}_{\rm A} + {\rm h.c.} ) ,
\end{equation} where h.c. denotes the hermitian conjugate. In the
same manner we define $ \hat{\Delta}_{\rm B }$. Then we form the
tensor product operator
\begin{equation}
\hat{\Delta}= \hat{\Delta}_{\rm A } \otimes \hat{\Delta}_{\rm B }
. \label{eq: delta general}
\end{equation}
We re-express this operator in terms of the system's sum and
difference phases, given in Eq. (\ref{eq: sums and differences})
above, as
\begin{eqnarray}
&&\hat{\Delta}(\varphi^{1,2}_{1,2 +},\varphi^{1,2}_{1,2 -}, \ldots
, \varphi^{1,2}_{N-1,N +},\varphi^{1,2}_{N-1,N -}\\\nonumber &&
,\varphi^{1,3}_{1,2 +}, \ldots , \varphi^{1,3}_{N-1,N -}, \ldots ,
\varphi^{M-1,M}_{1,2 +}, \ldots , \varphi^{M-1,M}_{N-1,N -}) .
\end{eqnarray}
This operator depends on $M(M-1)N(N-1)/2$ phase sums or
differences. We define, e.g.,
\begin{eqnarray}
\Gamma^{k,l}_{p,q +} &=& \frac{1}{2}  |\int_{2 \pi}
d\varphi^{k,l}_{p,q +} e^{i \varphi^{k,l}_{p,q +}}\\\nonumber &&
 {\rm Tr}(\oprho [ \hat{\Delta}(\varphi^{1,2}_{1,2
+},\varphi^{1,2}_{1,2 -}, \ldots ,\varphi^{M-1,M}_{N-1,N -})
\\\nonumber && + \hat{\Delta}(\varphi^{1,2}_{1,2
+}+\pi,\varphi^{1,2}_{1,2 -}+\pi, \ldots ,\varphi^{M-1,M}_{N-1,N
-}+\pi)  ] ) | . \label{eq:gammas 2}
\end{eqnarray}
Our final result, general to any bipartite system, is
\begin{eqnarray}\label{eq: final expression 2}
  \Gamma & = & ( 2\pi\\\nonumber
  &&{\cal N}_2 \sum_{k=1}^{M-1} \sum_{l=k+1}^{M} \sum_{p=1}^{N-1}
   \sum_{q=p+1}^{N} ||\Gamma^{k,l}_{p,q +}|-|\Gamma^{k,l}_{p,q -}||^2
  )^{1/2} \nonumber \\
  & = & \nonumber ( {\cal N}_2\sum_{k=1}^{M-1} \sum_{l=k+1}^{M} \sum_{p=1}^{N-1}
  \sum_{q=p+1}^{N} ||\oprho_{(k-1)N+p,(l-1)N+q}|\\ &&
  -|\oprho_{(k-1)N+q,(l-1)N+p}||^2
  )^{1/2} .
\label{eq: main result 2}
\end{eqnarray}
This is our main result, where the entanglement $\gmax$ of the
state $\oprho$ is understood to be the supremum of the equation
(\ref{eq: main result 2}) under all local unitary transformations.
For a separable state $\oprho = \oprho_{\rm A} \otimes \oprho_{\rm
B}$, the expression (\ref{eq: main result 2}), above, simplifies
to
\begin{eqnarray}
\g &=& ( {\cal N}_2\sum_{k=1}^{M-1} \sum_{l=k+1}^{M}
\sum_{p=1}^{N-1}
  \sum_{q=p+1}^{N} \\\nonumber &&
  ||\oprho_{{\rm A};kl}\oprho_{{\rm B};pq}|
  -|\oprho_{{\rm A};kl}\oprho_{{\rm B};qp}||^2
 )^{1/2} = 0 . \label{eq: separable state gives zero}
\end{eqnarray}
That is, our measure is identically zero for any separable state.
Hence, the expression can be used as a separability criteria for
any state, not only pure states. For a pure state where
\begin{equation}
\ket{\Psi}= \sum^{M}_{k=1}\sum^{N}_{l=1}\alpha_{kl}\ket{kl} ,
\end{equation}
we have
\begin{equation}
\Gamma = \left ( {\cal N}_2 \sum_{k=1}^{M-1} \sum_{l=k+1}^{M}
\sum_{p=1}^{N-1}
  \sum_{q=p+1}^{N} ||\alpha_{kp} \alpha_{lq}^*|-|\alpha_{kq} \alpha_{lp}^*||^2 \right
  )^{1/2} . \label{eq: separable state proof}
\end{equation}
We can compare this expression with the concurrence (or similar
measures such as the concurrence vector) for pure $M \otimes N$
bipartite states \cite{Audenaert,Rungta01,Albeverio,Gerjuoy}
\begin{equation}
C \propto \left ( \sum_{k=1}^{M-1} \sum_{l=k+1}^{M}
\sum_{p=1}^{N-1}
  \sum_{q=p+1}^{N} |\alpha_{kp} \alpha_{lq} - \alpha_{kq} \alpha_{lp}|^2 \right
  )^{1/2} . \label{eq: general concurrency}
\end{equation}
We conjecture that, properly normalized, $\gmax=C$ for any state. An extensive
numerical testing has always confirmed the hypothesis. We are
presently working on a strict mathematical proof of the
conjecture.

\section{Bipartite entanglement, the Bell states, and normalization}
\label{sec:examples}

From the previous section we conjecture that by measuring certain
of the density matrix coefficients, a pure state's entanglement
can be quantified. All the needed coefficients can be obtained by
projections on Bell states. E.g., the coefficient
$\oprho_{(k-1)N+p,(l-1)N+q}$ can be obtained by projection onto
the states $(\ket{kp} \pm \ket{lq})/\sqrt{2}$. We note that for an
$M \otimes N$ bipartite system, there exist $M(M-1)N(N-1)$ Bell
states. However, not all of those states are needed to measure the
bipartite state's entanglement. In analogy with our results for $2
\otimes 3$ systems, we conjecture that at most $M(M-1)N(N-1)/2$ of
the Bell states are needed, due to the degrees of freedom local
unitary transformations give us. For $M,N \geq 2$ this implies
that more than one Bell basis is needed, in general, to measure
the state's entanglement (we can note that when the product $M N$
is odd, complete Bell bases do not exist). At any rate, our
results demonstrate that for bipartite states, the Bell states
play a fundamental role in defining entanglement properties.

At first sight, the central role of the Bell states in
entanglement classification may seem obvious. However, perhaps
surprisingly, these states do not have the largest entanglement of
the bipartite states. For a Bell state, $\gmax = ({\cal
N}_2/2)^{1/2}$. However, the upper limit for $\gmax$ is reached
for, e.g., the state
\begin{equation}\label{eq: optimal gamma state}
  \frac{1}{\sqrt{K}}\sum^{K}_{k=1} \ket{kk} ,
\end{equation}
where $K$ is the smaller of $M$ and $N$. For this state we get
$\gmax={\cal N}_2^{1/2}$. When $M=N=K$, such states are,
e.g., manifested as the eigenstates of the relative-phase operator
\cite{Luis}, and they have been experimentally demonstrated
\cite{Trifonov}. The states also have a role in quantum
polarimetry \cite{Tsegaye}, and in quantum cryptography
\cite{Kulik}. In spite of this, they seen to have no direct role
as quantifying states of entanglement.

The observation above raises an important question: How do we
normalize $\gmax$ (or equivalently, the concurrence)? If we set
${\cal N}_2=1$ we get $0 \leq \gmax \leq 1$. However, a Bell
state in this Hilbert space will then have $\gmax =
(1/2)^{1/2}<1$. Another possibility is to set ${\cal N}_2=2$
rendering $\gmax =1$ for a Bell state and $0 \leq \gmax \leq
(2)^{1/2}$ in general. Ideally, one would like to find a
way to quantify entanglement so that one gets a quantitative
measure that corresponds to the state's utility in quantum
information tasks. Such a general quantification hierarchy lies
beyond the scope of this paper. Investigations of minimum
reversible entanglement generating sets
\cite{Bennett962,Bennett99,Acin01} have shown that even for
relatively simple Hilbert spaces, such as those housing tripartite
qubit systems, there exist two classes of states, W-states and
GHZ-states, that cannot be transformed into the other class by
local operations \cite{Dur,Galvao,Acin}. Hence, it is not clear to
us how an entanglement hierarchy should be defined.

\section{Discussion and conclusion}
\label{sec:conclusions}

In conclusion we have derived a quantitative measure of
entanglement for pure $M \otimes N$ bipartite quantum states,
based, essentially, on simple physical considerations. The measure
is positive, bounded, invariant to local unitary operations, and
it is shown that it equals zero for all separable states. Our
measure is less than, or equal to generalized concurrence for pure
bipartite state. We conjecture that it is actually always
proportional to the I-concurrence. In contrast to the latter, our
measure can be obtained by measurement in a direct fashion. To
this end, one projects the state onto Bell states and maximizes
the the sum of certain differences with respect to local unitary
transformations. The fact only projections onto Bell states are
needed suggests that these states are of particular significance
for all bipartite systems.

\begin{acknowledgments}
This work was supported by the Swedish Research Council (VR) and
the Swedish Foundation for Strategic Research (SSF).
\end{acknowledgments}

\end{document}